\documentclass[prd,11pt, a4paper,
nofootinbib,
preprintnumbers,
]{revtex4}
\usepackage{epsfig}
\usepackage{graphicx}
\usepackage[normal]{subfigure}
\usepackage{rotating}

\begin{document}
\title{Intermediate scales in non-supersymmetric SO(10) grand unified theories}
\date{September 14, 2009}
\author{Michal Malinsk\'{y}}\email{malinsky@kth.se} \thanks{\\In collaboration with Stefano Bertolini and Luca Di Luzio (SISSA \& INFN, Trieste). \\Talk given at {\it The 2009 Europhysics Conference on High Energy Physics,
		 July 16 - 22 2009
		 Krakow, Poland}}
\affiliation{Theoretical Particle Physics Group,
Department of Theoretical Physics,
Royal Technical Institute (KTH),
Roslagstullsbacken 21,
SE-106 91 Stockholm, Sweden.}
\begin{abstract}
It is often argued that in the class of non-supersymmetric $SO(10)$ grand unified theories there is barely any room  for reconciling the lower bound on the position of the GUT scale emerging from the proton decay searches and the lower limit on the absolute neutrino mass scale derived from the neutrino oscillation experiments with the gauge coupling unification constraints.
The recent two-loop reassessment of the gauge running provides the first complete picture of the situation, complementing the existing studies in several aspects. The improved analysis reveals a new room in the parametric space that could support a class of non-supersymmetric $SO(10)$ models potentially compatible with all current physical data, including constraints on the relevant Yukawa sector emerging from the quark and lepton masses and mixings. This, in turn, brings back the question of viability of some of the simplest non-supersymmetric GUT scenarios.
\vspace*{0ex}
\end{abstract}
\maketitle
\maketitle

\section{Introduction}
The grand unified theories (GUTs) provide one of the best motivated physics cases beyond the Standard Model (SM) paradigm. Recently, the information harvested from the neutrino oscillation experiments \cite{Schwetz:2008er} opened up another window to the fundamental GUT structures presumably underlying the flavour pattern of the SM, in particular the seesaw sector governing  the generation of the lepton masses and mixing. This triggered a considerable boost to the field by admitting a thorough reassessment of viability of the simplest GUT scenarios from a new perspective giving rise to a set of interesting recent results. A detailed understanding of the constraints emerging from the requirement of a high-energy gauge-coupling convergence represents a vital ingredient of any such efforts.

\section{Viability of the simplest GUTs}
Concerning the viability of various GUT models currently on the market there are several basic constraints to be taken into account. The present lower bound on the proton half-life 
$\tau_p(p\rightarrow e^+\pi^0) > 1.6\times 10^{33}$ years~\cite{Amsler:2008zz} yields a lower bound on the position of the unification scale $M_{GUT}\gtrsim 10^{15.4}$GeV. Second, the need to accommodate the ``atmospheric'' mass-squared difference in the neutrino sector via a variant of the seesaw mechanism implies the $B-L$ breaking scale should\footnote{The absolute neutrino mass scale constraint is naturally somewhat weaker than the one emerging from the proton decay. Moreover, the upper bound on $M_{B-L}$ is more rigid than the lower one because one can always look for cancellations in the seesaw formula if the ``naive'' estimate of the neutrino mass scale $m_{\nu}\sim v^{2}/M_{B-L}$ turns out too large; however, this is not the case for the opposite.} fall into the range $10^{12}{\rm GeV} \lesssim M_{B-L} \lesssim 10^{14}{\rm GeV} $. From this point of view one can briefly summarize the status of the simplest models based on the $SU(5)$ and $SO(10)$ gauge groups as follows:
\\{\bf SU(5) models:} It is well known that the minimal non-supersymmetric $SU(5)$ suffers from a multi-sigma defect in the gauge-coupling convergence. Although this problem can be alleviated in the SUSY context, the correlation between the flavour structures governing the $d=5$ proton decay and those controlling the quark and lepton masses and mixing is very tight in the minimal SUSY $SU(5)$ model, which can be reconciled with the current data only at the non-renormalizable level, c.f. \cite{Bajc:2002pg}.    
\\{\bf SO(10) models:} The $SO(10)$ models accommodate naturally the seesaw idea and provide much more room for intermediate scales. Nevertheless, it has been shown recently \cite{Aulakh:2005mw,Bertolini:2006pe} that in the minimal SUSY $SO(10)$ model \cite{Aulakh:2003kg} there is an inherent tension between the need for a relatively large atmospheric mass-squared difference in the neutrino sector (implying an upper bound on $M_{B-L}$ quite below $M_{GUT}$) and the requirements of the gauge coupling convergence (preferring $M_{B-L}$ close to $M_{GUT}$). This issue is absent in the non-SUSY case because then the separation between the GUT scale and (a set of) intermediate scales is vital for a successful gauge unification. 

\section{Intermediate scales in non-supersymmetric $SO(10)$ GUTs}
The constraints on the positions of intermediate scales in the non-SUSY $SO(10)$ GUTs imposed by the requirement of a successful gauge unification have been studied thoroughly in the past (see e.g. \cite{Chang:1984qr} for an initial assessment or \cite{Deshpande:1992em} for an update in view of the LEP data, and references therein). Nevertheless, with the new information on the flavour structure of the  lepton sector there are good prospects of putting the assessments of the viability of the simplest $SO(10)$ models on even a firmer ground. In general, in order to take the full advantage of the extra inputs one has to understand in detail the Higgs sector underpinning the flavour patterns of the model under consideration. Though simple in principle, this is often a rather complicated task in practice, but it is doubly rewarding as it provides another very important ingredient of an ultimate fully self-contained analysis of a particular setting -- the shape of the scalar sector thresholds populating the ``desert'' between the electroweak and the unification scales. This {\it a posteriori} justifies the two-loop approach used in some of the existing studies, despite the lack of an explicit Higgs sector treatment. 

\begin{figure}[ht]
\begin{center}
\includegraphics[width=6.5cm]{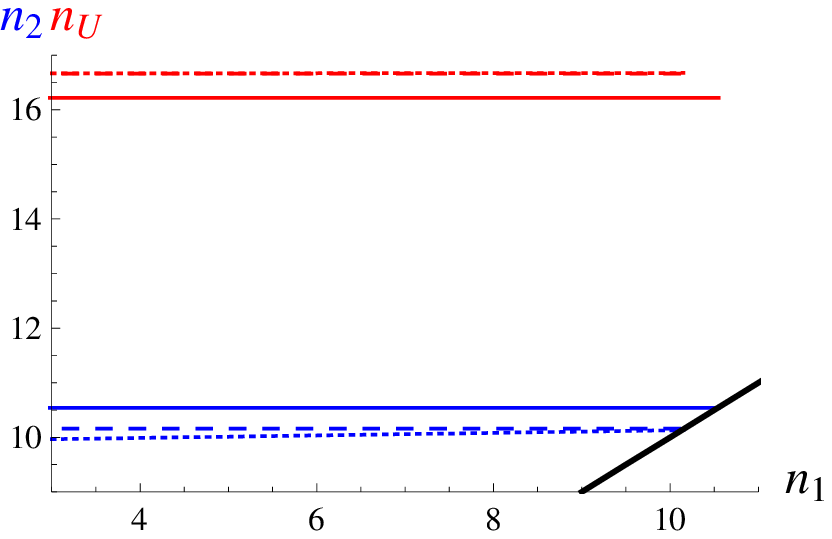}
\includegraphics[width=6.5cm]{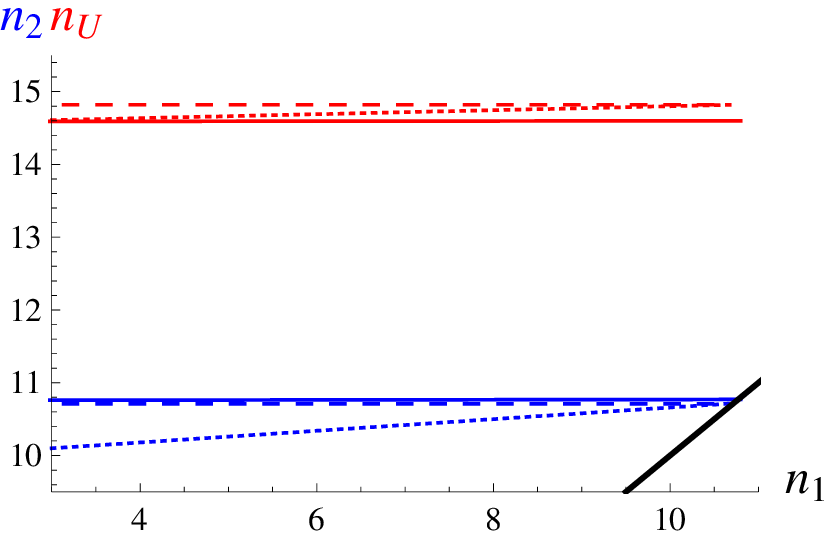}
\caption{\label{figs} The configurations of various intermediate scales corresponding to a pair of potentially interesting breaking chains ($SO(10)\stackrel{45}{\rightarrow} SU(3)_c\otimes SU(2)_L\otimes SU(2)_R\otimes U(1)_{B-L} \stackrel{45}{\rightarrow} SU(3)_c\otimes SU(2)_L\otimes U(1)_R\otimes U(1)_{B-L} \stackrel{16}{\rightarrow} SM$ on the left and $SO(10)\stackrel{45}{\rightarrow} SU(4)_C\otimes SU(2)_L\otimes U(1)_R\stackrel{45}{\rightarrow} SU(3)_c\otimes SU(2)_L\otimes U(1)_R\otimes U(1)_{B-L} \stackrel{126}{\rightarrow} SM$ on the right); $n_{U}$ and $n_{1}$ are defined in the text and $n_{2}\equiv \log_{10}(M_{2}/{\rm GeV})$ encodes the scale of the second breaking step. The solid lines depict the full two-loop results while the dashed (and dotted) lines correspond to the one-loop results with (and without) the $U(1)$-mixing effects taken into account. One can observe the generic tendency of the two-loop effects to rise the maximum $B-L$ scale while lowering the GUT scale. Remarkably, the the solid blue line in the diagram on the right is lifted by $\Delta n_{2}\sim 6$ (corresponding to a factor $\sim 10^{6}$ in maximum allowed $M_{B-L}$) in comparison with the previous assessments.} 
\end{center}
\end{figure}

From this perspective, the reappraisal \cite{Bertolini:2009qj} provides the first step towards such a complete viability analysis of the simplest non-SUSY $SO(10)$ models. Apart from fixing several shortcomings of the previous assessments\footnote{In particular,  we have corrected some of the numerical inconsistencies affecting substantially the results of the previous works even at the one-loop level (lifting a position of the $B-L$ scale by as much as almost six orders of magnitude in one specific case, c.f. the right panel in Figure \ref{figs}).} it extends the existing studies in several aspects. For instance, it provides the first fully consistent treatment of the so-called ``$U(1)$ mixing effects'' emerging in scenarios in which gauge group governing the dynamics below the GUT threshold contains more than a single $U(1)$ factor, with profound implications on the predictability of the relevant intermediate breaking scale, c.f. Figure \ref{figs}. Moreover, it does account for a need to bring a set of extra Higgs multiplets below the GUT scale if the Yukawa couplings are to be dominated by renormalizable couplings of the matter bilinears to $10_{H}\oplus \overline{126}_{H}$ in the Higgs sector\footnote{This mechanism is known to work only if the electroweak Higgs doublets contain  sizeable contributions from both these underlying multiplets; for this to be the case both components must be present at the scale the $SU(4)_{C}$ of Pati-Salam symmetry breaks down to $SU(3)_{c}\otimes U(1)_{B-L}$ above which the relevant doublets can not mix.} like in the minimal SUSY $SO(10)$ model. 

Remarkably enough, the new results affect considerably the status of some of the particularly attractive breaking chains, especially those featuring small representations governing the $SO(10)\to$ SM descent (thus often regarded to as minimal). In particular, we found considerable changes in the positions of the intermediate scales in the pair of chains \\ 
a) $SO(10)\stackrel{45}{\rightarrow} SU(3)_c\!\otimes\! SU(2)_L\!\otimes\! SU(2)_R\!\otimes\! U(1)_{B-L} \stackrel{45}{\rightarrow} SU(3)_c\!\otimes\! SU(2)_L\!\otimes\! U(1)_R\!\otimes\! U(1)_{B-L} \stackrel{16}{\rightarrow}$ SM,\\ 
b) $SO(10)\stackrel{45}{\rightarrow} SU(4)_C\otimes SU(2)_L\otimes U(1)_R \stackrel{45}{\rightarrow} SU(3)_c\otimes SU(2)_L\otimes U(1)_R\otimes U(1)_{B-L} \stackrel{126}{\rightarrow}$ SM, \\
as depicted in Figure \ref{figs}. In both cases, the two-loop effects tend to rise the (often too low) upper limit on the $B-L$ scale by almost an order of magnitude while lowering slightly the GUT-scale. In case a) one obtains $n_{U}\approx 16.2$ and $n_{1}\lesssim 10.8$ while in case b) $n_{U}\approx 14.8$ and $n_{1}\lesssim 11.0$ where $n_{U}\equiv \log_{10}(M_{GUT}/{\rm GeV})$ and $n_{1}\equiv \log_{10}(M_{B-L}/{\rm GeV})$. Concerning bounds on $n_{1}$ both configurations are compatible with the current experimental constraints even without a need to invoke a severe fine-tuning in the seesaw formulae. The value of $n_{U}$ in the latter case is not an issue either because the GUT-scale threshold effects can easily shift it by far more than the desired half order of magnitude. This, in turn, can be used as an extra physical requirement in the full-fledged analysis.

\section{Conclusions}
Flavour considerations provide strong constraints on the structure of the grand unified models. It is well known that the simplest supersymmetric $SU(5)$ and $SO(10)$ models suffer from severe shortcomings when it comes to a complete assessment including also the bounds emerging form the longevity of proton and/or from the need to reconcile the absolute light neutrino mass scale with the seesaw paradigm. On the non-SUSY side, the viability of certain chains within the class of simple $SO(10)$ models is still an open question deserving attention. The simplest strategy to address this issue requires a detailed two-loop understanding of the gauge coupling unification conundrum 
convoluted with a good apprehension of the Higgs sector. The analysis \cite{Bertolini:2009qj} is a further step towards this goal accounting for several inconsistencies plaguing the preceding assessments. The new results bring some of the previously disfavoured breaking chains back to life, with some of the improvements accounting for large changes  in the positions of the relevant intermediate scales. Remarkably, this affects namely the simplest chains relying on a minimal set of Higgs fields triggering the GUT symmetry breakdown, reopening the crucial question of their viability.



\begin{thebibliography}{1}

\bibitem{Schwetz:2008er}
  T.~Schwetz, M.~A.~Tortola and J.~W.~F.~Valle,
  New J.\ Phys.\  {\bf 10} (2008) 113011.

\bibitem{Amsler:2008zz}
Particle Data Group, C.~Amsler {\em et~al.},
\newblock Phys. Lett. {\bf B667}, 1 (2008).

\bibitem{Bajc:2002pg}
B.~Bajc, P.~Fileviez~Perez, and G.~Senjanovic,
[arXiv:hep-ph/0210374].

\bibitem{Aulakh:2005mw}
  C.~S.~Aulakh and S.~K.~Garg,
  Nucl.\ Phys.\  B {\bf 757} (2006) 47
  [arXiv:hep-ph/0512224].

\bibitem{Bertolini:2006pe}
  S.~Bertolini, T.~Schwetz and M.~Malinsky,
  Phys.\ Rev.\  D {\bf 73}, 115012 (2006).

\bibitem{Aulakh:2003kg}
  C.~S.~Aulakh, B.~Bajc, A.~Melfo, G.~Senjanovic and F.~Vissani,
  Phys.\ Lett.\  B {\bf 588}, 196 (2004).

\bibitem{Chang:1984qr}
  D.~Chang, R.~N.~Mohapatra, J.~Gipson, R.~E.~Marshak and M.~K.~Parida,
  Phys.\ Rev.\  D {\bf 31}, 1718 (1985).

\bibitem{Deshpande:1992em}
  N.~G.~Deshpande, E.~Keith and P.~B.~Pal,
  Phys.\ Rev.\  D {\bf 47}, 2892 (1993).

\bibitem{Bertolini:2009qj}
  S.~Bertolini, L.~Di Luzio and M.~Malinsky,
  Phys.\ Rev.\  D {\bf 80}, 015013 (2009).

\end{thebibliography}

\end{document}